\documentclass[conference,9pt]{IEEEtran}
\IEEEoverridecommandlockouts
\usepackage{cite}
\usepackage{amsmath,amssymb,amsfonts}
\usepackage{algorithmic}
\usepackage{graphicx}
\usepackage{textcomp}
\usepackage{xcolor}
\usepackage{hyperref}
\usepackage{booktabs}
\usepackage{multirow}
\usepackage{svg}
\usepackage[ruled,linesnumbered,vlined]{algorithm2e} 
\def\BibTeX{{\rm B\kern-.05em{\sc i\kern-.025em b}\kern-.08em
    T\kern-.1667em\lower.7ex\hbox{E}\kern-.125emX}}
\begin{document}

\newcommand{\papername}[1]{Orthrus}
\newcommand{\delayreduction}[1]{12.5\%}
\newcommand{\powerreduction}[1]{61.4\%}

% \SetAlgoSkip{\vspace{-10pt}}

% squish list
\newcommand{\squishlist}{
 \begin{list}{\scalebox{0.8}{$\bullet$}}
  {  \setlength{\itemsep}{0pt}
     \setlength{\parsep}{1pt}
     \setlength{\topsep}{2pt}
     \setlength{\partopsep}{0pt}
     \setlength{\leftmargin}{1.2em}
     \setlength{\labelwidth}{1em}
     \setlength{\labelsep}{0.5em} } }
\newcommand{\squishend}{
  \end{list}  }

\title{\papername{}: Dual-Loop Automated Framework for System-Technology Co-Optimization
\thanks{This work is supported in part by Beijing Natural Science Foundation (Grant No. L243001), National Natural Science Foundation of China (Grant No. 62032001, 62034007), National Key Research and Development Program of China (Grant No. 2023YFB4402204, 2021ZD0114702), and 111 Project (B18001).}}
% \title{Automated Framework for Feedback-Driven System-Technology Co-Optimization}
% \title{Automating Standard Cell Library Extension for Domain-Specific Computing Systems}

\author{
    \IEEEauthorblockN{
        Yi Ren\textsuperscript{1,2,\textdagger}, 
        Baokang Peng\textsuperscript{3,\textdagger},
        Chenhao Xue\textsuperscript{1},
        Kairong Guo\textsuperscript{1},
        Yukun Wang\textsuperscript{4}, 
        Guoyao Cheng\textsuperscript{3},\\
        Yibo Lin\textsuperscript{1,5,6},
        Lining Zhang\textsuperscript{3,5,*},
        Guangyu Sun\textsuperscript{1,5,6,*}
        \thanks{\textsuperscript{\textdagger}Co-first authors. \textsuperscript{*}Corresponding authors.}
    }
    \IEEEauthorblockA{
        \textsuperscript{1}\textit{School of Integrated Circuits}, 
        % \textsuperscript{1}\textit{School of Integrated Circuits, Peking University}, Beijing, China \\
        \textsuperscript{2}\textit{School of Software and Microelectronics, Peking University}, Beijing, China \\
         % Beijing, China \\
        \textsuperscript{3}\textit{School of Electronic and Computer Engineering, Peking University}, Shenzhen, China \\
        \textsuperscript{4}\textit{School of Electronics Engineering and Computer Science, Peking University}, Beijing, China \\
        \textsuperscript{5}\textit{Institute of Electronic Design Automation, Peking University}, Wuxi, China \\
        \textsuperscript{6}\textit{Beijing Advanced Innovation Center for Integrated Circuits}, Beijing, China \\
        % \{xch927027,gsun\}@pku.edu.cn
    }
    % \{yiren20, zjx\}@stu.pku.edu.cn, \{xch927027, yibolin, eelnzhang, gsun\}@pku.edu.cn, chenzhang.sjtu@sjtu.edu.cn, qxu@cse.cuhk.edu.hk
    % \{yiren20, baokangpeng\}@stu.pku.edu.cn, \{xch927027, yibolin, eelnzhang, gsun\}@pku.edu.cn, chenzhang.sjtu@sjtu.edu.cn, qxu@cse.cuhk.edu.hk\\
    \{yiren20, baokangpeng\}@stu.pku.edu.cn, \{eelnzhang, gsun\}@pku.edu.cn
    % \textsuperscript{*}Corresponding author
}

\newcommand{\xch}[1]{\textcolor{blue}{[XCH]: {#1}}}
\newcommand{\ry}[1]{\textcolor{green}{[RY]: {#1}}}
\newcommand{\bk}[1]{\textcolor{red}{[BK]: {#1}}}
% \and
% \and
% \and
% \and
% \and

\maketitle

\begin{abstract}
With the diminishing return from Moore's Law, system-technology co-optimization (STCO) has emerged as a promising approach to sustain the scaling trends in the VLSI industry. By bridging the gap between system requirements and technology innovations, STCO enables customized optimizations for application-driven system architectures. However, existing research lacks sufficient discussion on efficient STCO methodologies, particularly in addressing the information gap across design hierarchies and navigating the expansive cross-layer design space. To address these challenges, this paper presents \papername{}, a dual-loop automated framework that synergizes system-level and technology-level optimizations. At the system level, \papername{} employs a novel mechanism to prioritize the optimization of critical standard cells using system-level statistics. It also guides technology-level optimization via the normal directions of the Pareto frontier efficiently explored by Bayesian optimization. At the technology level, \papername{} leverages system-aware insights to optimize standard cell libraries. It employs a neural network-assisted enhanced differential evolution algorithm to efficiently optimize technology parameters. Experimental results on 7nm technology demonstrate that Orthrus achieves \delayreduction{} delay reduction at iso-power and \powerreduction{} power savings at iso-delay over the baseline approaches, establishing new Pareto frontiers in STCO. 
\end{abstract}


\section{Introduction}

Fabless-foundry business model serves as a cornerstone of modern VLSI industry, where fabless companies specialize in circuit design while foundries focus on manufacturing. The division of labor narrows the optimization objectives to specific domains, thereby facilitating decades of rapid industrial advancement. Unfortunately, the fabless-foundry model is now facing fundamental limitations. With design methodologies and associated automation tools reaching high maturity, further gains from design-level optimizations alone yield diminishing returns. Additionally, manufacturing process scaling is approaching its physical limits. To sustain the continued growth of the VLSI industry, deeper collaboration between fabless companies and foundries is becoming imperative, requiring a shift towards system-technology co-optimization (STCO) to unlock new performance and efficiency gains. According to Imec’s roadmap~\cite{biswas2024stco}, STCO is expected to play an increasingly vital role, particularly for application-driven system architectures.

\begin{figure}
    \centering
    \includegraphics[width=0.948\linewidth]{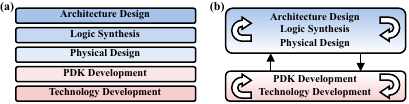}
    \caption{(a) Full-stack VLSI flow. (b) Our dual-loop STCO.}
    \label{fig:vlsi-stack}
    \vspace{-10pt}
\end{figure}

Conceptually, STCO aims to integrate multiple design hierarchies listed in Fig.~\ref{fig:vlsi-stack}(a), encompassing architecture design, logic synthesis, physical design, process design kit (PDK) development, and technology development. Each individual optimization level has been extensively investigated in prior research. At the architectural level, design space exploration (DSE) has been studied on various computing platforms, including CPU~\cite{bai2021boom,bai2023archexplorer}, AI accelerators~\cite{zhang2015optimizing,xiao2021hasco}, high-level synthesis~\cite{wang2021autosa,jia2021tensorlib}, and beyond. Similarly, numerous algorithms have been proposed to optimize the tunable parameters of logic synthesis tools and physical design tools~\cite{geng2022ppatuner,xie2020fist,liang2021flowtuner}. 
At the technology level, considerable research has focused on optimizing process parameters to enhance intrinsic device performance~\cite{iedmmos2,aimoocfet,tcasiifinfet} and standard cell performance~\cite{sdcop2,sdcop3}. In parallel, numerous studies have concentrated on improving the efficiency of standard cell characterization~\cite{sdcacc1,sdcacc2} and the generation of standard cell layouts~\cite{sdclayout1,sdclayout2}. 
% At the technology level, design and technology co-optimization (DTCO) has proven to yield significant benefits in recent years~\cite{DTCO_I,DTCO_II}. Current research focuses heavily on optimizing process parameters to enhance intrinsic device performance~\cite{iedmmos2,aimoocfet,tcasiifinfet}, as well as improving the performance of standard cells~\cite{sdcop2,sdcop3}. Meanwhile, considerable efforts have been devoted to improving the efficiency of standard cell characterization~\cite{sdcacc1,sdcacc2} and the generation of standard cell layouts~\cite{sdclayout1,sdclayout2}, to support PDK development effectively.

Unfortunately, despite extensive research on optimizations at individual design levels, the academic community lacks a systematic discussion of holistic optimization across the entire design hierarchy. This gap limits the translation of STCO's theoretical benefits to practical performance improvements.

On the one hand, a straightforward approach involves integrating multiple design hierarchies into a unified design space, where all relevant parameters are jointly optimized to maximize end-to-end quality-of-results (QoR). 
% Although this methodology shows promise for joint optimization of adjacent design levels~\cite{ma2018cross,liu2023boomerang,ren2025diffuse}, it suffers from fundamental scalability limitations when extended to the full STCO optimization chains: 
Although this methodology shows promise for joint optimization of adjacent design levels, such as system level DSE~\cite{ma2018cross,liu2023boomerang,ren2025diffuse} and design and technology co-optimization (DTCO)~\cite{DTCO_I,DTCO_II}, it suffers from fundamental scalability limitations when extended to the full STCO optimization chains: 
Firstly, a full-system evaluation using the complete design flow may take hours to days, rendering iterative optimization impractical; Secondly, the resulting high-dimensional design space exceeds the capabilities of existing DSE algorithms and cannot be efficiently navigated.

\begin{table*}[!t]
\centering
% \footnotesize
\caption{Cross-Layer Design Space of \papername{}.}
\resizebox{\linewidth}{!}{%
\begin{tabular}{c|c|c|c|c|c}
\toprule
ID  & Level & Parameter & Description & Candidate Values & Default Value \\
\midrule
1  & \multirow{2}{*}{Architecture} & \texttt{ct\_type} & compressor tree type & WT,DT & WT \\
2  & & \texttt{cpa\_type}                              & carry-propagate adder type & SK,KS,BK & SK \\
\midrule
3  & \multirow{4}{*}{\shortstack{Logic \\ Synthesis}}   & \texttt{clock\_period\_ns} & target clock period & range(0.4,1.0) & 0.5 \\
4  & & \texttt{syn\_generic\_effort}                    & generic synthesis effort & low,medium,high & medium \\
5  & & \texttt{syn\_map\_effort}                        & technology mapping effort & low,medium,high & high \\
6  & & \texttt{syn\_opt\_effort}                        & post-mapping optimization effort & none,low,medium,high & none \\
\midrule
7  & \multirow{5}{*}{\shortstack{Physical \\ Design}}   & \texttt{place\_utilization} & floorplan utilization ratio & range(0.5,0.9) & 0.8 \\
8  & & \texttt{place\_glb\_cong\_effort}                & effort for relieving congestion in global placement & auto,low,medium,high & auto \\
9  & & \texttt{place\_glb\_timing\_effort}              & effort for timing-driven global placement & medium,high & medium \\
10 & & \texttt{place\_glb\_clk\_power\_driven}          & enable clock tree power optimization in global placement & true,false & true \\
\midrule
11 & \multirow{6}{*}{Technology} & 
\texttt{phig\_n}  & nmos gate workfunction & range(4.302,4.312) & 4.307\\
12 & & \texttt{phig\_p}   & pmos gate workfunction & range(4.8631,4.8731) & 4.8681 \\
13 & & \texttt{hfin\_nm}  & height of fin & range(28,36) & 32 \\
14 & & \texttt{tfin\_nm}  & thickness of fin& range(5.8,7.2) & 6.5\\
15 & & \texttt{lg\_nm}    & horizontal length of the GATE layer & range(17,23) & 20\\
16 & & \texttt{lext\_nm}  & horizontal distance between the gate and the SDT layer & 4,5,6 & 5\\
17 & & \texttt{lct\_nm}   & horizontal length of the SDT layer & range(19,29) & 24\\ 
% 18 & & \texttt{num\_rows}        & range(1,3)\\
\bottomrule
\end{tabular}
} %
\label{tab:param_space}
\vspace{-10pt}
\end{table*}

On the other hand, we can retain the original design hierarchy and carefully coordinate their interactions to achieve overall benefits. However, establishing effective synergy across design levels remains a fundamental challenge. In the context of STCO, the primary challenge lies in bridging the gap between system-level performance, power, and area (PPA) metrics and technology innovations. Without visibility into standard cell criticality or well-defined guidance, optimization at the technology level cannot effectively mitigate system performance bottlenecks. While prior work has explored area reduction through merging common standard cell combinations~\cite{liang2022autocelllibx,fu2025temacle}, it remains an open problem to jointly address timing optimization, power reduction, and achieving intricate trade-offs among competing PPA objectives.

To address the above challenges, this paper introduces \textbf{\papername{}}, an automated framework to enable system-technology co-optimization, as shown in Fig.~\ref{fig:vlsi-stack}(b). \papername{} employs two synergistic optimization loops: 
The \textit{system loop} identifies Pareto-optimal parameters and collects data for directing technology optimization.
The \textit{technology loop} leverages system-level guidance to selectively optimize process parameters and standard cell layouts. 
% This optimization process combines neural networks with heuristic algorithms, enabling the efficient tuning of technology-related parameters to meet overall system performance requirements. 
% employs a novel transfer Gaussian process that harnesses the progressive characteristics of historically optimized libraries, enabling efficient expansion of the system-level Pareto frontier. 
The \textit{inter-loop direction} analyzes data from the system loop and guides the technology loop.

The main contributions of this paper are as follows: 

% \begin{itemize}
\squishlist
    \item We propose \papername{}, an automated STCO framework equipped with synergetic optimization loops.
    % \item We propose system-level feedback mechanism to enable targeted optimization of standard cell library.
    \item We propose a novel coordination mechanism that synergizes the system loop and technology loop by analyzing cell contributions, subcircuit frequencies, and PPA optimization directions.
    \item We propose a system optimization loop that leverages multi-objectives Bayesian optimization to efficiently identify the Pareto frontier while collecting data.
    \item We propose a technology optimization loop that leverages system-level guidance and employs a neural networks-assisted differential evolution algorithm to efficiently optimize technology parameters. 
    \item \papername{} achieves a 33.2\% PPA hypervolume improvement under advanced 7nm technology, delivering \delayreduction{} delay reduction at iso-power and \powerreduction{} power savings at iso-delay.
\squishend
% \end{itemize}

\begin{figure}[t]
    \centering
    \includegraphics[width=1\linewidth, trim = 65 73 65 73, clip]{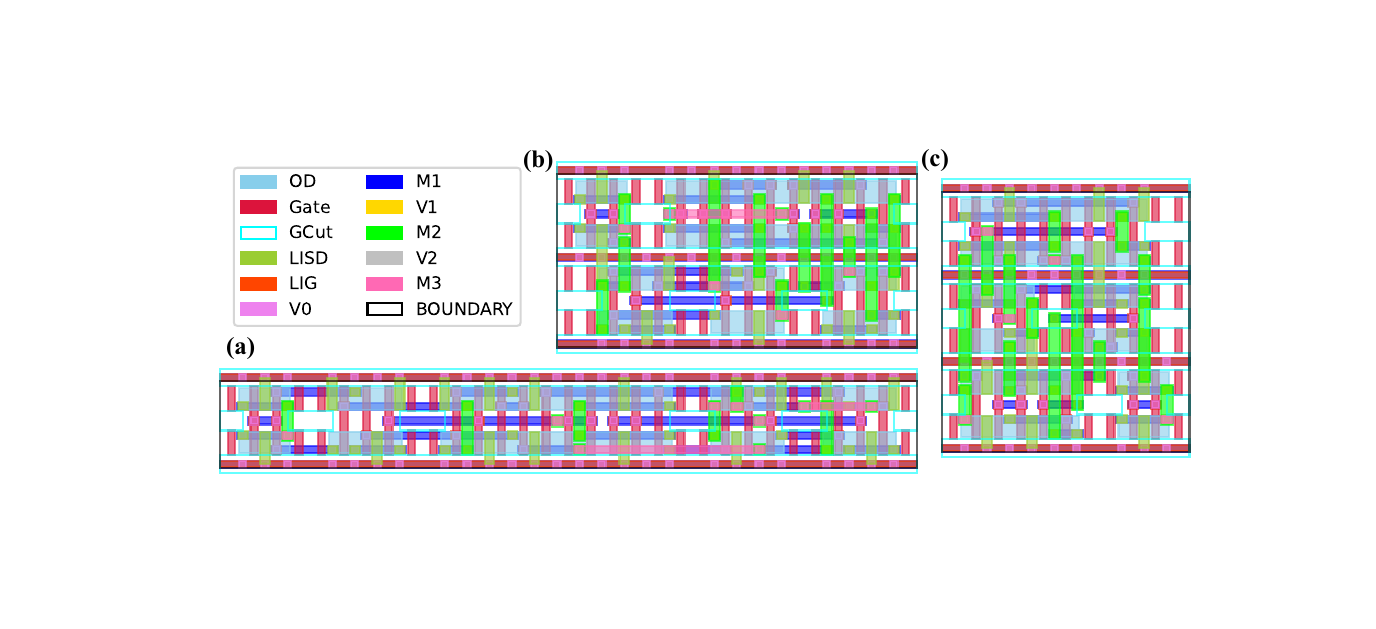}
    \caption{Layout of fused full-adder circuit (38 transistors): (a) Single-row configuration (\texttt{num\_rows} = 1) with 31 CPP width; (b) Two-row folded layout (\texttt{num\_rows} = 2) with 16 CPP width; (c) Three-row folded arrangement (\texttt{num\_rows} = 3) with 12 CPP width.}
    \label{fig:layout}
    \vspace{-10pt}
\end{figure}

The remainder of this paper is organized as follows:
Section~\ref{sec:preliminaries} provides preliminaries on graph matching and problem formulation.
Section~\ref{sec:methodology} details the \papername{} framework.
Section~\ref{sec:evaluation} presents the evaluation results.
Finally, Section~\ref{sec:conclusion} concludes the paper.

% 1. Largest chip e.g. Cerebrus
% 2. To handle the enormous design complexity, the VLSI industry divide and conqure
% 3..Specifically, as shown in fig.1, fabless is responsible for ..., whereas foundary is responsible 
% 4. Each design level has individual optimization objective and methodology, however it tends to fall into local optima
% 5. To fruther improve QoR, especially in the context of diminishing return from Moore's Law, we call for STCO. According to Imec's ...

\section{Preliminaries}
\label{sec:preliminaries}

\subsection{Graph Matching}
\label{sec:graph_matching}
Graph matching is a fundamental problem concerned with establishing correspondences or identifying structural similarities between graphs. It finds broad application in domains such as computer vision, pattern recognition, and circuit design. Classical approaches to graph matching include backtracking, depth-first search (DFS), and constraint-based pruning~\cite{yan2016short}.
% Graph matching is a fundamental problem concerned with establishing correspondences or identifying structural similarities between graphs. Algorithms developed for this purpose find widespread application across diverse domains, including computer vision, pattern recognition, and circuit design. Traditional graph matching approaches often encounter significant scalability challenges, which are typically solved with techniques such as backtracking, depth-first search (DFS), and constraint-based pruning~\cite{10.1145/2911996.2912035}.
In Electronic Design Automation (EDA), standard cell netlists are commonly represented as graphs, making graph matching techniques highly relevant. Subgraph isomorphism detection, a key aspect of graph matching, plays a crucial role in tasks such as Layout vs Schematic (LVS) verification and positioning of Integrated Clock Gating (ICG) cells~\cite{he2024efficient}. Specifically, a standard cell netlist $G$ is isomorphic to netlist $H$ if there exists a bijection mapping between their standard cell sets that preserves cell interconnections. 
% As depicted in Fig.~\ref{fig:Circuit_Graph}, 
Subgraph isomorphism detection aims to find all subgraphs within netlist $G$ that are isomorphic to an arbitrary query subgraph $H \subseteq G$. 

The discovery of isomorphic subgraphs enables various optimization opportunities, including standard cell merging. Prior works such as AutoCellLibX~\cite{liang2022autocelllibx} and TeMACLE~\cite{fu2025temacle} propose to merge frequent subcircuits for area reduction, utilizing the blank space within simple cells. However, these approaches focus solely on area reduction, neglecting cell delay and power consumption. As shown in Fig.~\ref{fig:layout}, \papername{} overcomes this limitation by incorporating multirow standard cell layout synthesis, which shortens the critical net length for improved delay and lower power dissipation~\cite{stdgen_kairong}.

\subsection{Problem Formulation}
\label{sec:problem formulation}

\papername{} employs a fully automated design flow that integrates EDA tools across multiple design levels. These tools offer a wide range of tunable parameters, creating an enormous cross-layer design space. TABLE~\ref{tab:param_space} summarizes the target design hierarchy and its associated parameters, detailed as follows:
% \begin{itemize}
\squishlist
    \item \textit{Architecture}: In \papername{}, we validate the efficacy of STCO methodology on application-driven system architectures, as customized design optimization for these architectures is expected to yield substantial practical benefits. Specifically, \papername{} targets the multiply-accumulator (MAC) arrays, a key component in AI accelerators that play a crucial role in determining the PPA of the entire system~\cite{sze2017efficient}. We employ an $8 \times 8$ systolic array with MAC units interconnected via pipeline registers. Each MAC unit incorporates a parallel multiplier architecture, comprising a partial product generator, a compressor tree, and a carry-propagate adder. We select compressor tree from Wallace Tree (WT) and Dadda Tree (DT), and carry-propagate adder from Sklansky adder (SK), Kogge-Stone adder (KS), and Brent-Kung adder (BK). An in-house RTL generator is developed to translate the MAC array configuration into Verilog HDL codes.
    \item \textit{Logic Synthesis}: The logic synthesis tool converts RTL implementations into standard cell netlists. \papername{} employs Cadence Genus for logic synthesis and adjusts the target frequency as well as synthesis efforts.
    \item \textit{Physical Design}: The physical design tool places and routes the standard cell netlists into a manufacturable circuit layout. In \papername{}, we employ Cadence Innovus for physical implementation. We mainly consider the design options at the global placement stage, since these options demonstrate a significant impact on PPA outcomes~\cite{liang2021flowtuner}.
    \item \textit{Technology}: To explore the parameters involved in technology optimization, \papername{} utilizes a customized ASAP7 open-source PDK~\cite{clark2016asap7} as an exemplary demonstration platform. We employ the calibrated model card from ASAP7 as the baseline model and adjust several model instance parameters. Besides, we adjusted the layout-related parameters of the standard cell and ensured that these adjustments satisfied the constant CPP requirement, as defined by the following equation:
    \begin{equation}
    \begin{aligned}
        CPP = L_g + 2*L_{ext}+ L_{ct}
    \end{aligned}
    \end{equation}
    For the cell layout, M1 and M3 are configured with 1D horizontal routing, while LISD and M2 use 1D vertical routing. 
    % The cell height is fixed at 6 tracks. To alleviate intra-cell routing congestion, we enable the multi-row layout for standard cells. While this may increase area due to the additional SDB, it can potentially lead to performance improvements. 
    For each standard cell, we validate the layout using Mentor Calibre to perform Design Rule Check (DRC), LVS, and Parasitic Extraction (PEX) checks, and extract the corresponding parasitics. Additionally, we use Cadence Liberate for delay and power characterization of the standard cells, generating the timing library (.lib). Cadence Abstract is employed to generate the physical library (.lef).
% \end{itemize}
\squishend

\textbf{Definition 1} (Tunable Parameter Design Space) 
\textit{A tunable parameter configuration $\mathbf{p}$ is defined as a combination of candidate values given in TABLE~\ref{tab:param_space}. The feature vector $\mathbf{p} = (\mathbf{p}_{arch}, \mathbf{p}_{ls}, \mathbf{p}_{pd}, \mathbf{p}_{tech})$ can be decomposed into multiple segments, each corresponds to the tunable parameters of a specific design level. The complete parameter design space $\mathcal{D}_{param}$ constitutes the set of all feasible parameter configurations.}

\begin{table}[!t]
\centering
\caption{Adopted Standard Cells from ASAP7 6T Library}
\label{tab:cell_lib}
\resizebox{\linewidth}{!}{%
\begin{tabular}{c|c|c}
\toprule
\textbf{Category} & \textbf{Standard Cells} & \textbf{Row Count} \\
\midrule
\multirow{4}{*}{Basic Cell} & AND2x2, AND2x4, AND3x1, NAND2x1, NAND2x2, NAND3x1, & \multirow{4}{*}{1}\\
 & OR2x2, OR2x4, OR3x1, NOR2x1, XNOR2x2, XOR2x2 & \\
 & INVx1, INVx2, INVx4, INVx8, BUFx2, BUFx4, BUFx8 & \\
 & MAJx1, MAJx2, AOI21x1, AO21x1, AO22x1, OA21x1, OA22x1 & \\
\midrule
Fused Cell & \textit{Extracted from frequent subcircuit patterns} & \{1,2,3\} \\
\bottomrule
\end{tabular}
} %
\vspace{-10pt}
\end{table}

\begin{figure*}
    \centering
    \includegraphics[width=0.9\linewidth]{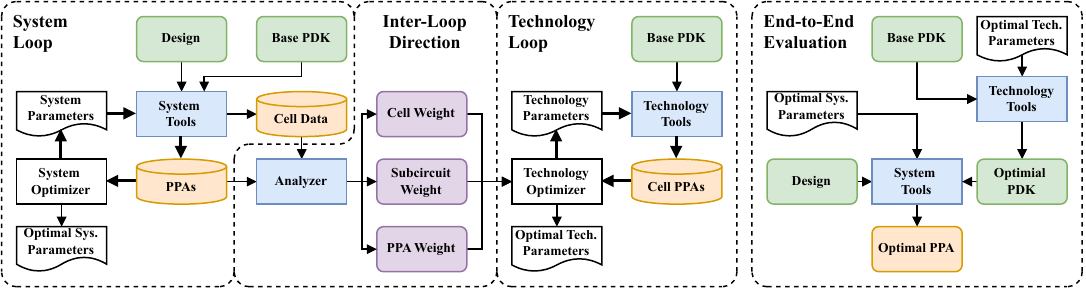}
    \caption{Overview of \papername{}. The system loop and technology loop search for optimal parameters at their respective levels. The inter-loop direction analyzes system loop data to guide technology loop optimization. End-to-end evaluation provides the PPA of optimal parameters.}
    \label{fig:overview}
    \vspace{-10pt}
\end{figure*}

In addition to adjusting the tunable parameters of EDA tools, \papername{} investigates the layout customization of individual standard cells. As illustrated in TABLE~\ref{tab:cell_lib}, \papername{} selects several fundamental standard cells from ASAP7 to establish the initial standard cell library, and extend the library by fusing subcircuits into new standard cells. We develop a C++ program for the automatic generation of multi-row standard cell layouts following~\cite{stdgen_kairong}, abbreviated as \texttt{StdGen}. In a nutshell, given the SPICE netlist of specific standard cells and target number of rows, \texttt{StdGen} systematically explores transistor placement while considering intra-cell routability, followed by SAT-based routing to ensure compliance with design rules. The generated layouts undergo DRC, LVS, PEX, and characterization, yielding optimized standard cells that replace the original ones in the subsequent design stages.

\textbf{Definition 2} (Cell Layout Design Space)
\textit{Given standard cell $c$, let $\mathcal{L}(c, R_c)$ denote the set of $c$'s feasible layouts whose row count falls in set $R_c$. For a standard cell library $\mathcal{C}$ given in TABLE~\ref{tab:cell_lib}, the cell layout design space $\mathcal{D}_{cell} = \prod_{c \in \mathcal{C}} \mathcal{L}(c, R_c)$ is defined as the Cartesian product of feasible layout sets for all standard cells in $\mathcal{C}$.}

Through joint optimization of the tunable parameters and standard cell layouts, \papername{} targets system-level improvements in PPA. Typically, these objectives are conflicting, where advancing one may degrade others. At the cell level, reducing the threshold voltage improves latency while increasing leakage power, and transistor width expansion enhances drive strength at the expense of a larger area. These trade-offs propagate to the system level, where performance gains incur either increased power dissipation or area overhead. In this context of multi-objective optimization, the optimal solutions form a Pareto frontier, where no PPA metrics can be further improved without deteriorating others. 
Since the true Pareto set cannot be obtained within limited trials in practical STCO scenarios, our objective is to advance the explored Pareto frontier, which is quantitatively measured by hypervolume improvement w.r.t. a reference point. Formally, our problem formulation and the related terminologies are defined as follows:  

\textbf{Definition 3} (Performance)
\textit{The performance is defined as the maximum attainable frequency of the MAC array, which is determined by the maximum delay of all timing paths.}

\textbf{Definition 4} (Power)
\textit{The power is defined as the average power dissipation when the MAC array operates at the maximum attainable frequency.}

\textbf{Definition 5} (Area)
\textit{The area is defined as the size of the floorplan in which the MAC array is placed and routed.}

\textbf{Definition 6} (Pareto Frontier)
\textit{Let objective vector $\mathbf{y}$ denote the PPA metrics. $\mathbf{y}$ is said to be Pareto-dominated by $\mathbf{y}'$ (denoted as $\mathbf{y} \preceq \mathbf{y}'$) if the following condition satisfies:
\begin{equation}
\begin{aligned}
    \forall i \in [1, 3], &\quad \mathbf{y}'[i] \leq \mathbf{y}[i]; \\
    \exists j \in [1, 3], &\quad \mathbf{y}'[j] < \mathbf{y}[j].
\end{aligned}
\end{equation}
Given a set of objective vectors $\mathcal{Y}$, its Pareto frontier is defined as a subset $\mathcal{Y}^* = \{\mathbf{y} | \mathbf{y} \not\preceq \mathbf{y}', \forall \mathbf{y}' \in \mathcal{Y}\}$}.

\textbf{Definition 7} (Hypervolume)
\textit{Given a set of objective vectors $\mathcal{Y}$ and a reference point $\mathbf{y}_{ref}$ that is strictly dominated by all $\mathbf{y} \in \mathcal{Y}$, the hypervolume (HV) is calculated as the Lebesgue measure of the dominated space:}
\begin{equation}
    HV(\mathcal{Y}^*, \mathbf{y}_{ref}) = \int_{\mathbb{R}^3} \mathbf{1} [\exists \mathbf{y}' \in \mathcal{Y}^*, \mathbf{y}' \preceq \mathbf{y} \preceq \mathbf{y}_{ref}] ~ d\mathbf{y}
\end{equation}

\textbf{Problem 1} (System-Technology Co-Optimization)
\textit{For subset $\mathcal{X} \subset \mathcal{D}_{param} \times \mathcal{D}_{cell}$ sampled from the joint design space of tunable parameters and standard cell layouts, its corresponding set of PPA metric $\mathcal{Y}$ can be obtained through the VLSI flow. Given limited invocation of the VLSI evaluation flow, the objective of \papername{} is to obtain $\mathcal{X}$ such that the hypervolume $HV(\mathcal{Y}, \mathbf{y}_{ref})$ can be maximized.}

\section{Methodology}
\label{sec:methodology}

\subsection{Framework Overview}
The overview of the \papername{} framework is shown in Fig.~\ref{fig:overview}.
First, the \textit{system loop} explores the system design space using Bayesian optimization, identifies Pareto-optimal parameters, and collects cell data (Section~\ref{sec:system-loop}).
Next, the \textit{inter-loop direction} analyzes this data through a novel mechanism to prioritize critical cells, suggest fusion candidates, and guide technology optimization (Section~\ref{sec:inter-loop}).
Then, the \textit{technology loop} optimizes technology parameters based on these system-aware insights via a neural network-assisted heuristic algorithm (Section~\ref{sec:tech-loop}).
Finally, \textit{end-to-end evaluation} provides the final PPA results of these optimal parameters.

\subsection{System Loop}
\label{sec:system-loop}

\begin{figure}
    \centering
    \includegraphics[width=1\linewidth]{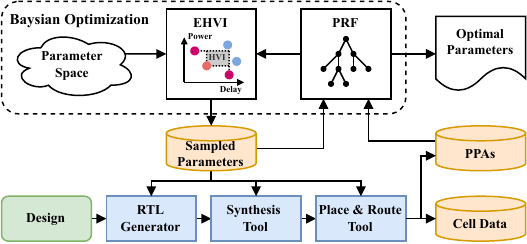}
    \caption{System loop uses EHVI and PRF for Bayesian optimization, along with an EDA toolchain to extract PPA metrics and cell data.}
    \label{fig:system-loop}
    \vspace{-10pt}
\end{figure}

% \ry{Finish me.}
The proposed system loop framework, as shown in Fig.~\ref{fig:system-loop}, aims to identify Pareto-optimal parameters and collect cell data for directing technology optimization. 
Bayesian Optimization (BO)~\cite{bo2016shahriari} is adopted to navigate high-dimensional parameter spaces efficiently, overcoming the computational infeasibility of brute-force sampling in multi-objective optimization. 
By synergizing surrogate modeling with automated design toolchains, the framework balances exploration of under-sampled regions and exploitation of known high-performance solutions, while efficiently correlating system parameters with PPA.

\begin{algorithm}[t]
\caption{Bayesian Optimization}
\label{algo:bo}
\SetKwInOut{Input}{Input}
\SetKwInOut{Output}{Output}
\Input{Parameter Space $\mathcal{D}$, Maximum Iteration $t_{max}$}
\Output{Pareto frontier $\mathcal{Y}^*$ and corresponding Parero set $\mathcal{X}^*$}
Initialize $\mathcal{X}_0$ via random sampling from $\mathcal{D}$\;
Evaluate $\mathcal{Y}_0$ via toolchain\;
\For{$t \gets 1$ \KwTo $t_{max}$}{
    Train surrogate model $M$ on $(\mathcal{X}_{t-1}, \mathcal{Y}_{t-1})$\;
    Select $\mathbf{x}_t = \arg\max \alpha(\mathbf{x})$\;
    Evaluate $\mathbf{y}_t$ via toolchain\;
    Update $\mathcal{X}_t = \mathcal{X}_{t-1} \cup \{\mathbf{x}_t\}$, $\mathcal{Y}_t = \mathcal{Y}_{t-1} \cup \{\mathbf{y}_t\}$\;
}
\Return Pareto frontier $\mathcal{Y}^*$ and corresponding Parero set $\mathcal{X}^*$
\end{algorithm}

\textbf{Bayesian Optimization.} The BO algorithm iteratively refines parameter selections using a surrogate model $M$ and an acquisition function $\alpha(\cdot)$. Let $\mathcal{X}_t = \{\mathbf{x}_i\}_{i=1}^t$ and $\mathcal{Y}_t = \{\mathbf{y}_i\}_{i=1}^t$ denote the evaluated parameters $\mathbf{x}_i$ and their objective vectors $\mathbf{y}_i$. At each iteration, the surrogate model approximates the posterior distribution of $\mathbf{y}$, and the acquisition function $\alpha(\mathbf{x})$ prioritizes candidate points. The pseudocode is shown in Algorithm~\ref{algo:bo}.

\textbf{Initialization and Acquisition Function.} The framework initializes with random sampling to ensure spatial coverage of the parameter space. Subsequent iterations employ Expected Hypervolume Improvement (EHVI)~\cite{ehvi2020daulton} as the acquisition function to maximize hypervolume gains on the Pareto frontier. For a candidate $\mathbf{x}$, EHVI quantifies the expected improvement over the current Pareto frontier $\mathcal{Y}^*$:  
\begin{equation}  
\text{EHVI}(\mathbf{x}) = \mathbb{E}\left[\max\left(0, HV(\mathcal{Y}^* \cup \mathbf{y}(\mathbf{x})) - HV(\mathcal{Y}^*)\right)\right],  
\end{equation}  
where $HV(\cdot)$ is the simplified notion for $HV(\cdot,\mathbf{y}_{ref})$, and the expectation integrates over the surrogate model’s predictive distribution.  

\textbf{Surrogate Model.} To avoid incompatibility or high computational complexity in Bayesian optimization, Probabilistic Random Forest (PRF)~\cite{prf2011hutter} serves as the surrogate model, extending standard random forests by outputting Gaussian distributions for each objective. For an ensemble of $B$ regression trees, PRF predicts the mean $\mu(\mathbf{x}) = \frac{1}{B}\sum_{b=1}^B \mu_b(\mathbf{x})$ and variance $\sigma^2(\mathbf{x}) = \frac{1}{B}\sum_{b=1}^B (\mu_b(\mathbf{x}) - \mu(\mathbf{x}))^2$ for each objective. This probabilistic formulation enables uncertainty-aware EHVI computation, crucial for balancing exploration-exploitation trade-offs.  

\textbf{EDA toolchain.} The toolchain integrates three stages: RTL Generator synthesizes parameterized hardware descriptions, Synthesis Tool maps RTL to gate-level netlists, and Place \& Route Tool generates physical layouts and reports the PPA $\mathbf{y}$ for BO. Cell data—extracted from the final netlist—includes the timing of critical paths and power/area of each cell, forming the database for inter-loop analysis. This closed-loop system automates parameter-to-PPA translation, enabling system-aware technology optimization.

\subsection{Inter-Loop Direction}
\label{sec:inter-loop}

\begin{figure}
    \centering
    \includegraphics[width=1\linewidth]{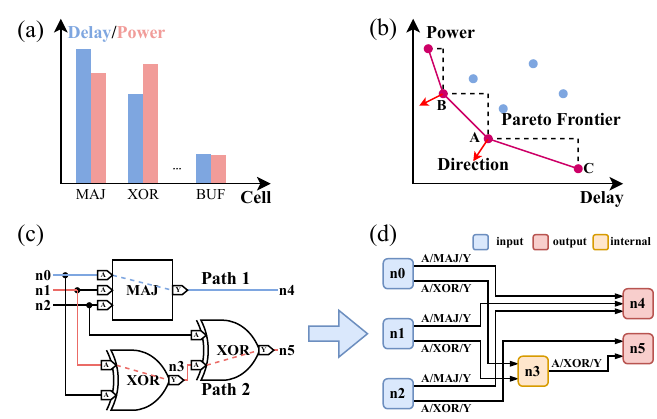}
    \caption{Inter-loop analysis. (a) Per-cell delay/power contributions. (b) PPA directions on Pareto frontier. (c) Per-cell critial timing path. (d) Standard cell netlists modeled with net-centric directed acyclic graph (DAG).}
    \label{fig:interloop}
    \vspace{-10pt}
\end{figure}

This section presents the coordination mechanism that synergizes system loop and technology loop. 
As shown in Fig.~\ref{fig:overview}, \papername{} analyzes the post-routing netlists alongside corresponding system-level PPA metrics to guide technology optimization.
The resulting inter-loop direction consists of three key components: (1) the PPA contribution of each standard cell type; (2) the occurrence frequency of cell combinations; (3) the optimization direction for specific system-level parameter configurations.

In the following subsections, we will introduce the details of each type of inter-loop direction.

\textbf{Cell Contribution Analysis.}
As shown in Fig.~\ref{fig:interloop}(a), we quantify the contribution of each cell to system performance and power consumption, which enables prioritized optimization on critical cells. Our study focuses on power and timing impact, considering that cell area remains unchanged after process parameter tuning and cell layout exploration.

% The analysis of system-level PPA contribution of the standard cells can provide critical insights for technology-level optimization. First, since process parameter optimization in the technology loop affects all standard cells, identifying key contributors enables focused optimization of standard cells with substantial system-level PPA impact. Second, cell-level importance analysis facilitates prioritized layout optimization of critical standard cell. Given that cell area remains fixed during technology-level optimization, our analysis concentrates on standard cell contribution to system-level timing and power objectives.

The power contribution of standard cell $c$ can be derived from its aggregate power consumption divided by the total system power. Formally, given a post-routing standard cell netlist $\mathcal{G}$, the power contribution of $c$ is calculated by:
\begin{equation}
    w_c^{power} = \frac{\sum_{g \in \mathcal{G}} power(g)\cdot \mathbf{1}[Type(g) = c]}{power(\mathcal{G})}
\end{equation}

The timing contribution of standard cell $c$ is determined through path-based analysis, where cells appearing more frequently on critical timing paths are considered to have a greater impact on system-level timing performance. The rationale stems from the observation that such cells are either essential components of timing-critical functional modules or favored by synthesis tools to mitigate timing violations. In either case, optimizing these standard cells can effectively enhance overall performance. To quantify the timing contribution of cell type $c$, we first compute the timing contribution of each individual cell instance $g$ (Equation~\eqref{eq:cell_inst_timing_score}), then derive the aggregated contribution for $c$ by averaging the score across all corresponding instances (Equation~\eqref{eq:cell_timing_score}).
\begin{equation}
    w_g^{delay} = \exp(\lambda \cdot \max \{ delay(p) \mid g \in p, p \in \mathcal{P}\})
\label{eq:cell_inst_timing_score}
\end{equation}
\begin{equation}
    w_c^{delay} = \frac{\sum_{g \in \mathcal{G}} w_g^{delay}\cdot \mathbf{1}[Type(g) = c]}{\sum_{g \in \mathcal{G}} w_g^{delay}}
\label{eq:cell_timing_score}
\end{equation}
In Equation~\eqref{eq:cell_inst_timing_score}, a timing path $p$ is a signal route across the cell netlist, as illustrated in Fig.~\ref{fig:interloop}(c). The set of all timing paths is denoted as $\mathcal{P}$. The timing path delays are obtained using the static timing analysis engine within Innovus. Conceptually, the contribution of standard cell instance $g$ diminishes exponentially with larger quantity and greater criticality of competing timing paths. The hyperparameter $\lambda$ is introduced to modulate the weighting mechanism's sensitivity to competitive effects.

\textbf{Subcircuit Analysis.}
In \papername{}, we propose to synthesize multirow standard cells for frequently occurring cell combinations, aiming to improve system-level PPA. A customized subgraph isomorphism detection algorithm is employed to identify these common cell combinations.

% \begin{figure}
%     \centering
%     \includegraphics[width=1\linewidth]{figure/Building_Graph.pdf}
%     \caption{The standard cell netlists are modeled with directed acyclic graph (DAG), where nodes represent nets and edges represent the cells and corresponding pins between nets.}
%     \label{fig:Building_Graph}
% \end{figure}

Before diving into algorithmic details, we first describe the graph construction process. As illustrated in Fig.~\ref{fig:interloop}(d), we employ a net-centric representation to model the standard cell netlists. Specifically, a combinatorial standard cell netlist can be represented as a directed acyclic graph (DAG). The vertices correspond to inter-cell nets and are categorized into three types (\texttt{Input/Output/Internal}). Edges represent intra-cell connections and are annotated with the cell type and associated I/O pins. For netlists that contain sequential elements (e.g. registers and latches), we partition the design into multiple combinatorial subcircuits and apply the graph matching algorithm independently.

\begin{algorithm}[t]
\small

\caption{Frequent Subcircuit Mining}
\label{algo:subgraph}
\SetKwInOut{Input}{Input}
\SetKwInOut{Output}{Output}

\newcommand\mycommfont[1]{\footnotesize\ttfamily\textcolor{blue}{#1}}
\SetCommentSty{mycommfont}

\Input{Netlist $G=(V_G,E_G)$, max depth $d_{max}$, max output number $o_{max}$, max input number $i_{max}$}
\Output{Subgraph occurence count $M: \Sigma^* \mapsto \mathbb{N}$}

$M \leftarrow \{s \mapsto 0 \mid \forall s \in \Sigma^* \}$\;
\ForAll{$V_H \subseteq V_G, |V_H| \leq o_{max}$}{
    \tcc{Explore subcircuits using DFS.}
    $V_P \leftarrow \{v \in V_G \mid \exists u \in H, v \text{~can reach~} u \text{~in~} G\}$\;
    $P \leftarrow G[V_P]$ \tcp{Induced subgraph}
    $\mathcal{S}_H \leftarrow \texttt{DFS}(P, d_{max}, i_{max})$\;
    \tcc{Count subcircuit patterns.}
    \ForAll{$S \in \mathcal{S}_H$}{
        $s \leftarrow \texttt{CanonicalRepr}(S)$\;
        $M[s] \leftarrow M[s] + 1$\;
    }
}
\Return $M$\;

\end{algorithm}

% \begin{algorithm}
% \label{algo:subgraph}
% \small

% \caption{Frequent Subcircuit Mining}
% \SetKwInOut{Input}{Input}
% \SetKwInOut{Output}{Output}

% \newcommand\mycommfont[1]{\footnotesize\ttfamily\textcolor{blue}{#1}}
% \SetCommentSty{mycommfont}

% \Input{Circuit $C$, max depth $d$, subcircuit number $N$}
% \Output{List of frequent subcircuits}

% \SetKwFunction{Explore}{ExploreCells}
% \SetKwProg{Fn}{Function}{:}{}
% \Fn{\Explore{$gate\_id$, $depth$, $current\_pins$, $visited$, $reprs$}}{
%     Generate subgraph $S$ from $current\_pins$\;
%     \If{$S$ inputs $\leq 4$ and outputs $\leq 2$}{
%         $canon\_repr \gets$ GetCanonicalRepr($S$)\;
%         \If{$canon\_repr \neq \emptyset$}{
%             Add $canon\_repr$ to $reprs$\;
%             Update $representative\_subgraphs$\;
%         }
%         \If{$depth < d$}{
%             Find predecessor gates of $gate\_id$\;
%             \ForAll{predecessor subsets}{
%                 Update $current\_pins$ and $visited$\;
%                 \Explore{new state}\;
%             }
%         }
%     }
% }

% \tcc{Process Gate Roots}
% \ForAll{gate pairs $(g_1, g_2)$}{
%     Initialize combined pins\;
%     \Explore from both gates\;
%     Update $direct\_cnt$\;
% }

% \tcc{Hierarchical Counting}
% \ForAll{$repr_i \in sorted\_reprs$}{
%     \ForAll{$repr_j \in sorted\_reprs$}{
%         \If{$repr_i$ is specialization of $repr_j$}{
%             $total\_cnt[repr_j] \mathrel{+}=direct\_cnt[repr_i]$\;
%         }
%     }
% }

% Return top $N$ sorted by $total\_counts$\;
% \end{algorithm}

Algorithm~\ref{algo:subgraph} outlines the frequent subcircuit mining process. In a nutshell, for netlist $G = (V_G, E_G)$ represented with net-centric DAG, the algorithm systematically explores all connected subcircuits within a bounded size and records the occurrence frequency of subcircuit patterns. To ensure tractability, we impose the constraints that candidate subcircuit $S \subseteq G$ must have input net count $\texttt{NumIn}(S) \leq i_{max}$, output net count $\texttt{NumOut}(S) \leq o_{max}$, and logic depth $\texttt{Depth}(S) \leq d_{max}$. In practice, we set $i_{max}=4, o_{max}=2, d_{max}=3$, respectively. 
% We first select a net set $V_H$. Then, we employ depth-first search (DFS) to enumerate all valid subcircuits that have $V_H$ as sink ports and satisfy the above constraints, yielding a subcircuit set $\mathcal{S}_H$.
Each traversed subcircuit $S$ is hashed into a unique key using an established colored DAG hashing method~\cite{helbling2020directed}. We record the frequency of its corresponding subcircuit pattern via bucket counting.

\textbf{PPA Direction Analysis.}
As illustrated in Fig.~\ref{fig:interloop}(b), the direction of technology optimization is derived from the geometric properties of the Pareto frontier identified in the system loop.  
Given the computational overhead of iteratively evaluating the technology toolchain, we formulate a single-objective optimization for the technology loop by weighting the PPA metrics.  
Specifically, for each Pareto-optimal point, we calculate the normal vector to the local Pareto frontier using Singular Value Decomposition (SVD) on its $ k $-nearest neighbors $\mathcal{N}_k$. This vector defines the trade-off sensitivity between delay and power, expressed as $\mathbf{v}_2^\top = [-W_{delay}, -W_{power}]$, the last row of $ V^\top $:  
\begin{equation}  
V^\top = \begin{bmatrix} \mathbf{v}_1^\top \\ \mathbf{v}_2^\top \end{bmatrix}, \quad \mathcal{N}_k = U \Sigma V^\top.
\label{eq:ppa-direction}
\end{equation}  
Additionally, we flip the direction if it is far from the origin.  
As shown in Fig.~\ref{fig:interloop}(c), with this direction, the optimization balances power and delay at the balanced point $ A $, while pushing the delay to the limit at the low-delay point $ B $. 

\begin{figure}
    \centering
    \includegraphics[width=1\linewidth]{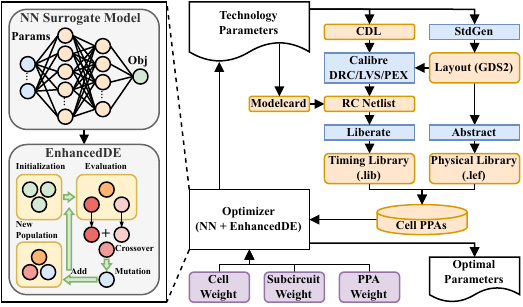}
    \caption{Technology loop diagram, utilizing an EDA toolchain to extract PPA metrics for each standard cell, accepting weight inputs, employing neural network as a surrogate model, and using EnhancedDE as the optimizer to output the optimal parameters.}
    \label{fig:techloop}
    \vspace{-10pt}
\end{figure}

\begin{figure}[!t]
    \centering
    \includegraphics[width=1\linewidth]{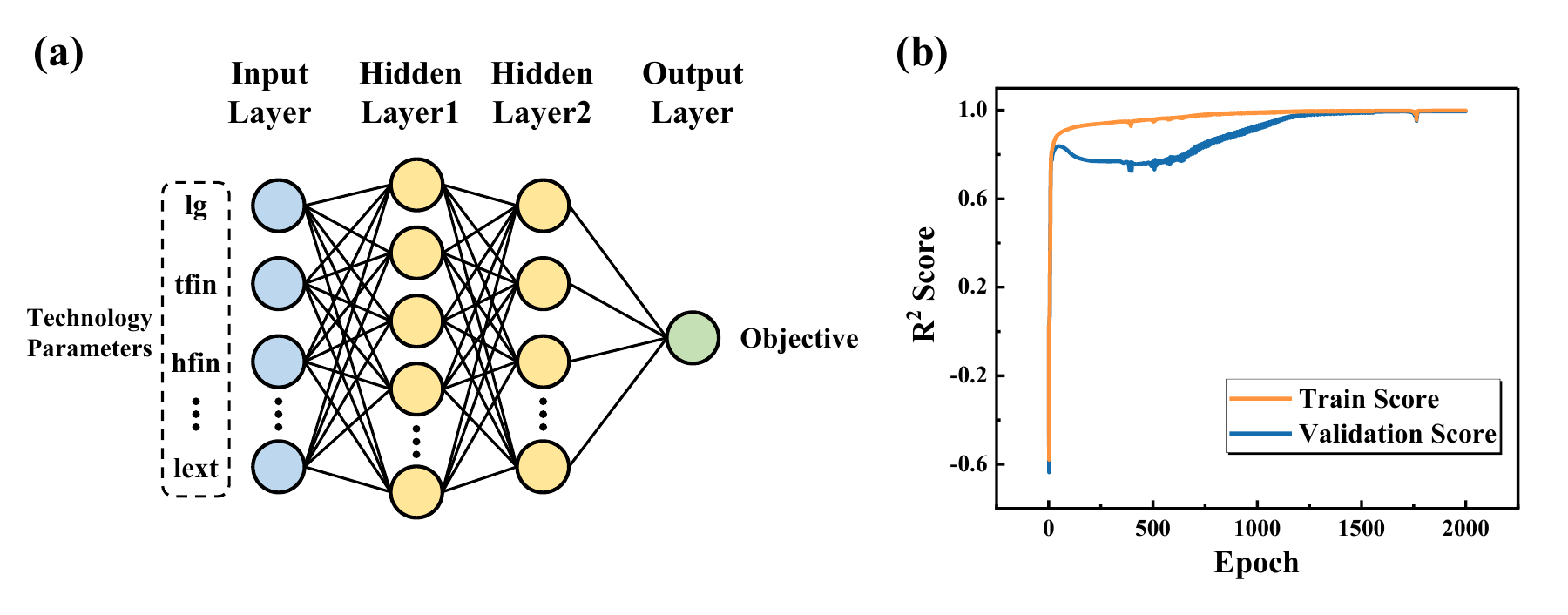}
    \vspace{-20pt}
    \caption{(a) Schematic diagram of the neural network surrogate model; \\(b) $R^{2}$ scores for the training and validation sets as a function of epochs during the neural network training process.}
    \label{fig:trainloop}
    \vspace{-10pt}
\end{figure}

\subsection{Technology Loop}
\label{sec:tech-loop}

\begin{algorithm}[t]
\small
\caption{Neural Network-Assisted EnhancedDE for Technology Optimization.}
\label{algo:dtco}

\SetKwInOut{Input}{Input}
\SetKwInOut{Output}{Output}
\newcommand\mycommfont[1]{\footnotesize\ttfamily\textcolor{blue}{#1}}
\SetCommentSty{mycommfont}

\Input{Initial samples $N$, NN model training epochs $E$, maximum iteration for enhancedDE $I_{max}$, DE generation $n_{gen}$, population size $s_{pop}$, top size $s_{top}$, mutation factor $MF$, penalty factor $PF$, penalty threshold $PT$, crossover probability $CR$, historical data $\mathbf{X}_{his}$, $\mathbf{C}_{his}$}
\Output{Optimal parameters $\mathbf{x}^*$}

\tcc{Phase 1: Initial Sampling}
$\mathbf{X}_{init} \leftarrow \texttt{LHS}(N)$ \tcp{Latin Hypercube Sampling}
$\mathbf{C}_{init} \leftarrow \texttt{CellSimulate}(\mathcal{P}_{init})$ \tcp{standard cell simulation}
$(\mathbf{Y}_{init}, \mathbf{Y}_{his}) \leftarrow \texttt{PPACalculation}(\mathbf{C}_{init}, \mathbf{C}_{his})$\;
$\mathbf{D} \leftarrow (\mathbf{X}_{init}, \mathbf{C}_{init}, \mathbf{Y}_{init})$ \tcp{Initial dataset}
$\mathbf{D} \leftarrow (\mathbf{D}.\mathbf{X} \cup \mathbf{X}_{his}, \mathbf{D}.\mathbf{C} \cup \mathbf{C}_{his}, \mathbf{D}.\mathbf{Y} \cup \mathbf{Y}_{his})$\;

\For{$i = 1 \rightarrow I_{max}$}{
    \tcc{Phase 2: Surrogate Model Construction}
    $\texttt{NNSurrogate}(\cdot) \leftarrow \texttt{TrainMLP}(\mathbf{D}, \text{epochs}=E)$\;
    \tcc{Phase 3: EnhancedDE Optimization}
    $\mathbf{P} \leftarrow \texttt{InitPopulation}(s_{pop}, \mathbf{D}.\mathbf{X})$\;
    $\mathbf{T} \leftarrow \emptyset$ \tcp{Elite solution archive}

    \For{$j = 1 \rightarrow n_{gen}$}{
        \ForAll{$x \in P$}{
            \If{$\texttt{\upshape rand}() < 0.2$ and $|\mathbf{T}| <2$}{
                $a \leftarrow \texttt{RandomSelect}(\mathbf{P} \setminus \{x\})$\;
                $b, c \leftarrow \texttt{RandomSelect}(\mathbf{T}, 2)$\;
            }
            \Else{
                $a, b, c \leftarrow \texttt{RandomSelect}(\mathbf{P} \setminus \{x\}, 3)$\;
            }
            $m \leftarrow a + MF\times(b- c)$ \tcp{mutant}
            $t \leftarrow \texttt{CrossOver}(x, m, CR)$ \tcp{trial}
            $f_{base} \leftarrow \texttt{NNSurrogate}(t)$\;
            $d_{min} \leftarrow \texttt{MinDistance}(t, \mathbf{T} \cup \mathbf{P})$\;
            $f_{penalized} \leftarrow f_{base} + PF \times \texttt{max}(0, PT - d_{min})$\;
            \If{$f_{penalized} < \texttt{\upshape fitness}(x)$}{
                $\mathbf{P}.\texttt{update}(x, t, f_{penalized})$\;
            }
            $\mathbf{T} \leftarrow \texttt{UpdateElites}(\mathbf{T} \cup \{t\}, PT)$
        }
    }

    \tcc{Phase 4: Design Verification \& Update}
    $\mathbf{X}_{candidates} \leftarrow \texttt{SelectDiverse}(\mathbf{T}, s_{top})$\;
    $\mathbf{C}_{true} \leftarrow \texttt{CellSimulate}(\mathbf{X}_{candidates})$\;
    $\mathbf{Y}_{true} \leftarrow \texttt{PPACalculation}(\mathbf{C}_{true})$\;
    $\mathbf{D} \leftarrow (\mathbf{D}.\mathbf{X} \cup \mathbf{X}_{candidates}, \mathbf{D}.\mathbf{C} \cup \mathbf{C}_{true}, \mathbf{D}.\mathbf{Y} \cup \mathbf{Y}_{true})$\;
}

\tcc{Phase 5: Final Parameter Extraction}
$k^* \leftarrow \texttt{argmin}(\mathbf{D}.\mathbf{Y})$\;
$\mathbf{x}^* \leftarrow \mathbf{D}.\mathbf{X}[k^*]$\;
\Return $\mathbf{x}^*$\;

\end{algorithm}

The proposed technology loop framework, as illustrated in Fig.~\ref{fig:techloop}, aims to optimize technology parameters for system-level performance. The framework consists of three primary steps: \texttt{CellSimulate}, \texttt{PPADirected}, and \texttt{Optimizer}. These steps enable efficient simulation and optimization of technology parameters for system-level design.

The \texttt{CellSimulate} function takes as input the parameters at the Technology level from TABLE~\ref{tab:param_space} and outputs the PPA of each standard cell. The simulation process can be broken down into the following stages:
\squishlist
    \item \textit{Parameter Adjustment}: The input parameters are used to adjust the circuit netlist, model card, and \texttt{StdGen} configuration. This ensures that the simulation aligns with the target technology parameters.
    \item \textit{Layout Generation}: The \texttt{StdGen} function takes the configuration parameters and generates the corresponding standard cell layouts. The generated layouts are then validated through DRC, LVS, and PEX to verify the correctness of the layout and extract the parasitic netlists.
    \item \textit{Library Generation}: Using the validated layouts, the physical library (.lef) is created. Additionally, the extracted parasitic netlists, along with the model card, are used to generate the timing library (.lib), which contains the necessary delay and power characterizations for each standard cell. This results in the power, delay, and area information for each standard cell, which corresponds to the output of the \texttt{CellSimulate} function.
    % \item Cell PPA Calculation: The PPA of each standard cell is computed by comparing the physical and timing libraries generated in Step 3 with the reference library. This comparison provides the PPA values for each standard cell, which are then used as the output of the CellSimulate function.
\squishend

% As described in Section~\ref{sec:problem formulation}, Definition 2, fusing frequently occurring subcircuits into standard cells can help improve the system-level PPA. However, due to the large number of transistors in the fused cells, it is necessary to consider num\_rows as a parameter to be optimized. To reduce the dimensionality of the model, we optimize only the num\_rows parameter for each fused cell, rather than considering the num\_rows of every individual standard cell. The optimization results, including the impact of cell fusing on the overall system, will be discussed in detail in Section IV.B.

The \texttt{PPACalculation} function uses PPA weights along with the delay and power of each cell and returns the corresponding weighted PPA objective $y$ for the technology loop. $y$ is computed according to the formula in Equations~\eqref{eq:weighted_delay}-\eqref{eq:weighted_obj}. First, the normalized cell delay and power are weighted by their respective contribution (see cell contribution analysis in Section~\ref{sec:inter-loop}). Normalization references original ASAP7 cells for those from the initial library, and the initial single-row version for the fused cells. Next, system-level direction weights aggregate delay and power metrics to evaluate the optimization objective (see PPA direction analysis in Section~\ref{sec:inter-loop}).
% \begin{equation}
%     \begin{aligned}
%         & Delay(\mathbf{C}) = \sum\limits_{c \in \mathbf{C}} {w_{c}^{delay}}  \times norm\_delay(c) \\
%         & Power(\mathbf{C}) = \sum\limits_{c \in \mathbf{C}} {w_{c}^{power}}  \times norm\_power(c) \\
%         & \mathbf{Y} = {W_{delay}} \times Delay(\mathbf{C}) + {W_{power}} \times Power(\mathbf{C}) \\
%     \end{aligned}
%     \label{equ:powercal}
% \end{equation}
\begin{equation}
Delay(\mathbf{C}) = \sum\limits_{c \in \mathbf{C}} {w_{c}^{delay}} \times norm_{delay}(c)
\label{eq:weighted_delay}
\end{equation}
\begin{equation}
Power(\mathbf{C}) = \sum\limits_{c \in \mathbf{C}} {w_{c}^{power}} \times norm_{power}(c)
\label{eq:weighted_power}
\end{equation}
\begin{equation}
y = {W_{delay}} \times Delay(\mathbf{C}) + {W_{power}} \times Power(\mathbf{C})
\label{eq:weighted_obj}
\end{equation}

The \texttt{Optimizer} step, detailed in Algorithm~\ref{algo:dtco}, involves leveraging a neural network as a surrogate model to reduce reliance on the time-consuming \texttt{CellSimulate} function, while employing an Enhanced Differential Evolution (EnhancedDE) algorithm to optimize the technology parameters. The specific process is outlined as follows:

% \begin{figure}
%     \centering
%     \includegraphics[width=1\linewidth, trim = 65 73 65 73, clip]{figure/FA_mr_v2.pdf}
%     \caption{Layout of fused full-adder circuit (38 transistors): (a) Single-row configuration (num\_rows = 1) with 31 CPP width; (b) Two-row folded layout (num\_rows = 2) with 16 CPP width; (c) Three-row folded arrangement (num\_rows= 3) with 12 CPP width.}
%     \label{fig:layout}
% \end{figure}

\squishlist

\item \textit{Initialization Sampling:}
The initialization sampling employs Latin Hypercube Sampling (LHS) for broad parameter space coverage, with historical values from the previous technology loop integrated to expand the neural network’s training dataset. This helps avoid the selection of optimal solutions that overlap with historical solutions, improving the diversity of the sampling process.

\begin{figure}[!t]
    \centering
    % \begin{minipage}[b]{\linewidth}
    \includegraphics[width=1\linewidth]{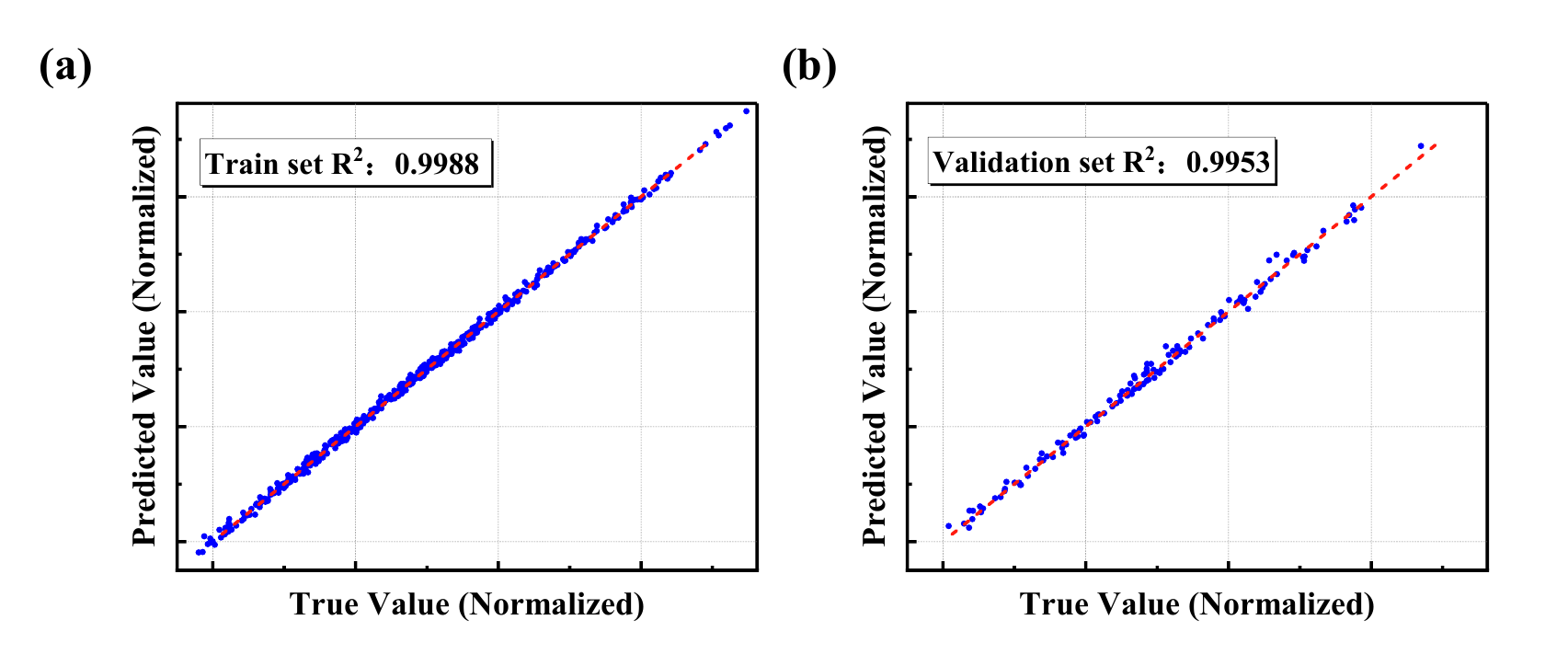}
    % \end{minipage}
    \vspace{-20pt}
    \caption{Comparison between surrogate model predictions and ground truth values: (a) Training set performance showing strong agreement ($R^{2}$ = 99.88\%), with data points closely distributed along the diagonal trend line; (b) Test set performance (20\% holdout) demonstrating model generalizability, maintaining good correlation ($R^{2}$ = 99.53\%). Dashed lines represent perfect prediction ($y = x$).}
    \label{fig:nnvalidation}
    \vspace{-15pt}
\end{figure}

\item \textit{Neural Network Surrogate Model:}
We employ a fully connected neural network (as shown in Fig.~\ref{fig:trainloop}(a)) to predict the objective $y$ based on normalized input parameters, as listed in TABLE~\ref{tab:param_space}. The model’s output is the predicted objective value $\hat{y}$. The loss function is defined as follows:
\begin{equation}
    Loss = MSE(\hat{y} - y) = \frac{1}{n} \sum_{i=1}^{n} (\hat{y_{i}}  -y)^{2} 
    \label{equ:Loss}
\end{equation}
To prevent overfitting, we use batch processing for the training set and incorporate L2 regularization. The training and testing dataset variations with respect to the number of epochs are shown in Fig.~\ref{fig:trainloop}(b), and the validation results for the training and testing datasets are shown in Fig.~\ref{fig:nnvalidation}. The $R^{2}$ value for the training dataset is 99.88\%, for the testing dataset is 99.53\%.
The results demonstrate that the model effectively predicts the target values with a high degree of accuracy.

\item \textit{Differential Evolution Optimization:}
The \texttt{Optimizer} uses an EnhancedDE algorithm, which builds on traditional differential evolution and adds a distance penalty to improve the diversity of the solution. While the neural network surrogate model accelerates the evaluation process, the EnhancedDE algorithm utilizes the population search strategy to perform extensive exploration of the parameter space, ultimately identifying the optimal set of solutions. To further ensure solution diversity, a distance penalty is introduced. This penalty evaluates the Euclidean distance between candidate solutions and existing solutions, checking if the distance exceeds a predefined penalty threshold ($PT$). If the candidate solution fails to meet the distance requirement, a penalty is applied, discouraging the selection of similar solutions and promoting diversity in the final batch of optimal solutions.
Upon completion of one iteration, the optimal solutions returned are evaluated using the \texttt{CellSimulate} and \texttt{PPADirected} functions to obtain the true target value, $\mathbf{Y}_{true}$, which then updates the dataset $\mathbf{D}$. Once the iteration reaches the predefined $I_{max}$, the optimal parameter vector $\mathbf{x}^*$ corresponding to the best target value is returned.
% Although introducing the distance penalty increases the complexity of the algorithm, it does not significantly affect the overall runtime, as the CellSimulate function remains the primary bottleneck. The distance penalty enhances the diversity of the final solution set, leading to better optimization results.

\squishend

% As discussed in Section~\ref{sec:problem formulation}, fusing frequently occurring subcircuits into standard cells can help improve the system-level PPA. 
The \texttt{Optimizer} is responsible for adjusting both technology parameters $\mathbf{p}_{tech}$ and cell-specific hyperparameters \texttt{num\_rows}. Since the latter presents more complexity, we hereby make further elaborations on this process.
Given a base standard cell library derived from the original ASAP7 library, \papername{} selects the $N_{ext}$ most frequent subcircuit patterns for cell fusion (detailed in subcircuit analysis from Section~\ref{sec:inter-loop}). The selected subcircuits are assigned to an initial cell row count $\texttt{num\_rows}=1$ and are incorporated into the library extension. Whereas the hyperparameter of the remaining subcircuit patterns is permanently assigned to 0, meaning that these subcircuits will not be fused and added to the library. As discussed in Section~\ref{sec:graph_matching}, \texttt{num\_rows} serves as a key hyperparameter for fused cells with numerous transistors, which balances area compactness and critical path length. In the subsequent invocation of technology loop, we adjust \texttt{num\_rows} of the selected fused cells between 1 and 3 to explore this trade-off. For all standard cells adopted from the initial ASAP7 library, \texttt{num\_rows} is fixed to 1 to reduce the complexity of the surrogate model.

\begin{figure*}
    \centering
    \begin{minipage}[b]{0.3\textwidth}
    \includegraphics[width=1\linewidth]{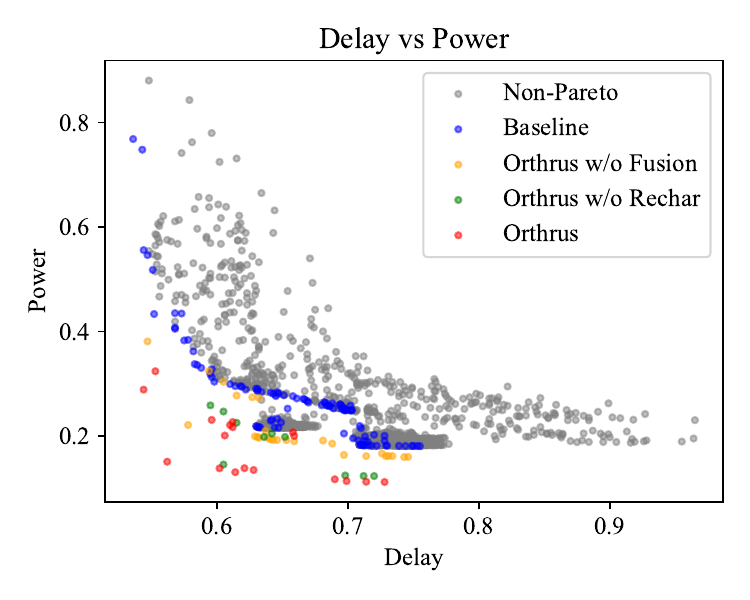}
    \end{minipage}
    % \hfill
    \begin{minipage}[b]{0.3\textwidth}
    \includegraphics[width=1\linewidth]{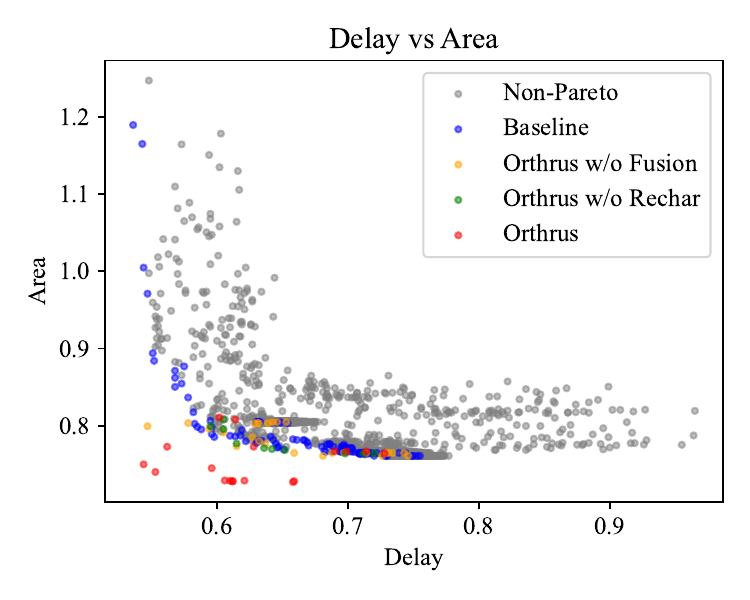}
    \end{minipage}
    % \hfill
    \begin{minipage}[b]{0.3\textwidth}
    \includegraphics[width=1\linewidth]{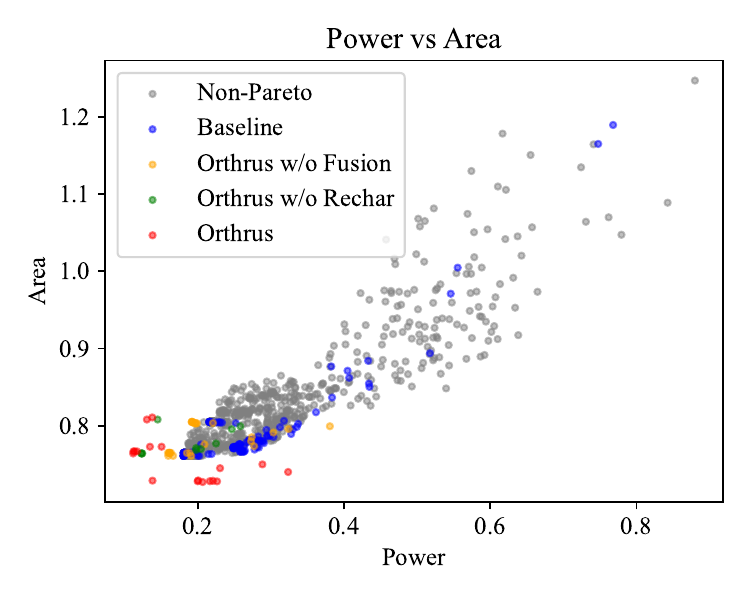}
    \end{minipage}
    \vspace{-10pt}
    \caption{Normalized Pareto frontier of baseline, \papername{} without subcircuit fusion, \papername{} without standard cell recharacterization, and \papername{}. The reference point for computing hypervolume is $(1,1,1)$.}
    \label{fig:pareto-front}
    \vspace{-10pt}
\end{figure*}

\begin{figure*}[t]
    \centering
    \begin{minipage}{0.18\linewidth}
    \centering
    \includegraphics[height=2.5cm]{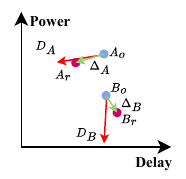}
    % \subcaption{figure caption}
    \end{minipage}%
    % \hfill
    \begin{minipage}{0.8\linewidth}
    \centering
\resizebox{!}{1cm}{%
% \begin{tabular}{l|c|c|c|c}
% \toprule
% \textbf{Metric \& Param.} & $A_o$ & $B_o$ & $A_r$ & $B_r$ \\ 
% \midrule
% Power    & 0.313 & 0.145 & 0.289 & 0.131 \\
% Delay    & 0.622 & 0.605 & 0.544 & 0.614 \\
% \midrule
% phig\_n  & \multicolumn{2}{c|}{4.307}  & 4.302  & 4.312  \\ 
% phig\_p  & \multicolumn{2}{c|}{4.8681} & 4.8682 & 4.8681 \\ 
% hfin\_nm & \multicolumn{2}{c|}{32}     & 36     & 28     \\ 
% tfin\_nm & \multicolumn{2}{c|}{6.5}    & 7.1    & 5.8    \\ 
% lg\_nm   & \multicolumn{2}{c|}{20}     & 17     & 18     \\ 
% lext\_nm & \multicolumn{2}{c|}{5}      & 6      & 6      \\ 
% lct\_nm  & \multicolumn{2}{c|}{24}     & 25     & 24     \\ 
% num\_rows & \multicolumn{2}{c|}{1}     & 3      & 1      \\ 
% \midrule
% \textbf{Cosine} & \multicolumn{2}{c|}{-}     & 0.989      & 0.822      \\ 
% \bottomrule
% \end{tabular}
\begin{tabular}{l|cc|cccccccc|c}
\toprule
\textbf{Category} & \textbf{Power} & \textbf{Delay} & \textbf{phig\_n} & \textbf{phig\_p} & \textbf{hfin\_nm} & \textbf{tfin\_nm} & \textbf{lg\_nm} & \textbf{lext\_nm} & \textbf{lct\_nm} & \textbf{num\_rows} & \textbf{Cosine} \\
\midrule
$A_o$ & 0.313 & 0.622 & 4.307 & 4.8681 & 32 & 6.5 & 20 & 5 & 24 & 1 & \multirow{2}{*}{0.989} \\
$A_r$ & 0.289 & 0.544 & 4.302 & 4.8683 & 36 & 7.1 & 17 & 6 & 25 & 3 &  \\
\midrule
$B_o$ & 0.145 & 0.605 & 4.307 & 4.8681 & 32 & 6.5 & 20 & 5 & 24 & 1 & \multirow{2}{*}{0.822} \\
$B_r$ & 0.131 & 0.614 & 4.312 & 4.8680 & 28 & 5.8 & 18 & 6 & 24 & 1 &  \\
\bottomrule
\end{tabular}
} %
    % \subcaption{table caption}    
    \end{minipage}
    % \hfill
    \caption{Directional alignment between the optimized parameter vector $D$ and the actual optimization trajectory $\Delta$ (subfigure). The corresponding power and delay metrics for each category, along with the technology level parameters and cosine similarity (subtable).}
    \label{fig:techvalidation}
    \vspace{-10pt}
\end{figure*}

\section{Evaluation}
\label{sec:evaluation}

\subsection{Setup}

\textbf{Platform.}
The automated STCO framework runs on a Linux-based platform with an Intel(R) Xeon(R) Gold 6342 CPU @ 2.80GHz and 1536 GiB of memory. 
% The technology loop is run on a Linux-based platform with an Intel(R) Xeon(R) Gold 5218R CPU @ 2.10GHz, and 125 GiB of memory. 
Cadence Genus 19.12-s121\_1 and Cadence Innovus v21.14-s109\_1 are used to synthesize, place, and route every sampled design. Cadence Liberate 19.2.1.215, Cadence Abstract 6.1.8, Cadence Spectre 18.1.0, and Mentor Calibre v2019.3\_15.11 are used for characterizing the timing library, physical library, circuit simulation, layout verification, and parasitic extraction.
The Bayesian Optimization in the system loop is implemented via \texttt{ParallelOptimizer} in OpenBox 0.8.4~\cite{li2021openbox} with default settings.

\textbf{Hyperparameters.}
We set the sensitivity parameter $\lambda = 10$ in Equation (\ref{eq:cell_inst_timing_score}) for computing cell timing contribution. We choose $k=2$ neighbors in Equation (\ref{eq:ppa-direction}) for finding the normal vector.
The neural network surrogate model in the technology loop adopts an architecture with two hidden layers (16 and 8 neurons respectively) utilizing sigmoid activation functions. An initial learning rate of 0.02 is configured with the Adam optimizer for parameter updates. The training epoch $E$ for the neural network is set to 1500 epochs to ensure convergence. In the optimization algorithm, the parameters are set as follows: $PT$ = 0.1, $PF$ = 1e3, $MF$ = 0.8, $CR$ = 0.9, $s_{pop}$ = 100, $n_{gen}$ = 20, $s_{top}$ = 5, and $I_{max}$ = 2.

\textbf{Baseline.}
To demonstrate the efficacy of technology optimization, the baseline approach only adjusts system-level parameters $(\mathbf{p}_{arch}, \mathbf{p}_{ls}, \mathbf{p}_{pd})$. We use the default technology parameters from TABLE~\ref{tab:param_space} and the basic standard cells from TABLE~\ref{tab:cell_lib}.

\subsection{Result Analysis}

\textbf{Pareto frontier and Hypervolume.}
We identify two key techniques for significant PPA improvement: (1) Standard cell recharacterization (Rechar), which adjusts technology parameters $\mathbf{p}_{tech}$ and \texttt{StdGen} hyperparameter \texttt{num\_rows}; (2) Subcircuit fusion (Fusion), which merges common subcircuit patterns into new standard cells. We ablate their individual and combined contributions to expanding PPA Pareto frontiers, as demonstrated in Fig.~\ref{fig:pareto-front} and TABLE~\ref{tab:hypervolume}. Without Fusion, adjusting only $\mathbf{p}_{tech}$ achieves a 6.5\% hypervolume improvement over the baseline. Without Rechar, fusing subcircuits into single-row standard cells yields a 7.7\% hypervolume improvement over the baseline. When these techniques are combined, we observe significant reductions in delay and power along with moderate area savings, resulting in a substantial hypervolume improvement of 33.2\%.
% As demonstrated in Fig.~\ref{fig:pareto-front} and TABLE~\ref{tab:hypervolume}, \papername{} achieves significant PPA improvements through standard cell recharacterization\footnote{Select the optimized technology parameters from TABLE~\ref{tab:param_space}, and when applying subcircuit fusion, also adjust \texttt{num\_rows}.} (Rechar) and subcircuit fusion\footnote{Only fuse the subcircuits while keeping all technology parameters at their default values.} (Fusion). The normalized Pareto frontiers (reference point is $(1,1,1)$) clearly show progressive optimization across all PPA dimensions. The application of either Rechar or Fusion yields reductions in both delay and power while maintaining comparable area utilization, consequently improving HV by 6.5\% and 7.7\%. When these techniques are combined, we observe significant reductions in delay and power along with moderate area savings, resulting in a substantial HV improvement of 33.2\%.
We further measured the optimization results of individual metrics while maintaining others constant (allowing a tolerance of 1e-3). 
Due to the observed power-area correlation ($r=0.88$), we focus on delay-power tradeoffs: achieving \powerreduction{} power savings at iso-delay and \delayreduction{} delay reduction at iso-power conditions.

\begin{table}[!t]
\centering
% \footnotesize
\caption{Hypervolume of Each Method}
\resizebox{\linewidth}{!}{%
\begin{tabular}{c||c|c|c|c}
\toprule
\textbf{Method}                                  & \textbf{Baseline}  & \textbf{\papername{} w/o Fusion} & \textbf{\papername{} w/o Rechar} & \textbf{\papername{}} \\
\midrule
\multirow{2}{*}{HV ($\times 10^{-2}$)}  & 8.055     & 8.582  & 8.679  & 10.727 \\
                                        & -         & +6.5\% & +7.7\% & 
 +33.2\% \\
\bottomrule
\end{tabular}
} %
\vspace{-10pt}
\label{tab:hypervolume}
\end{table}

\begin{figure}
    \centering
    \includegraphics[width=1\linewidth]{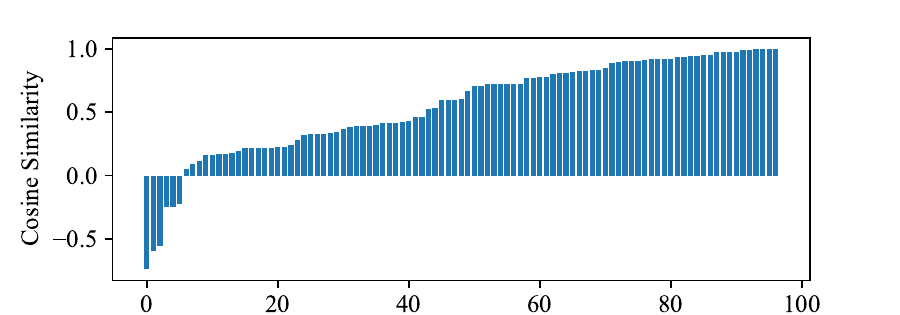}
    \caption{The sorted cosine similarity between the PPA optimization direction and the actual recharacterization path.}
    \label{fig:cosine-similarity}
    \vspace{-10pt}
\end{figure}

\textbf{Effectiveness of Inter-Loop Direction.}
To validate the effectiveness of inter-loop direction, we quantify the cosine similarity between the optimization direction and the actual Rechar path. As depicted in Fig.~\ref{fig:cosine-similarity}, which plots the \textit{sorted} cosine similarity across all optimization points, the vast majority of values are positive (clustering near or reaching 1.0). This strong alignment confirms that the optimization path adheres closely to the inter-loop direction.
Additionally, we evaluated the optimization results using naive cell weighting (i.e., treating all cells as equally important). The final hypervolume of $9.443\times10^{-2}$ represents a 12.0\% reduction compared to the results of \papername{}. This outcome demonstrates the effectiveness of our prioritized cell weighting approach during optimization.
% Naive standard cell weight: 9.443 (-12.0\%).

\textbf{Subcircuit Fusion.}
Statistical analysis based on the methodology introduced in Section~\ref{sec:inter-loop} reveals that Full Adders (FAs) and Half Adders (HAs) account for the majority of the delay (53.1\%), power (65.2\%), and area (75.7\%) overhead. Consequently, we specifically optimize these two subcircuits through fusion techniques.

\textbf{Case study.}
To further investigate the effectiveness of our proposed method, we present two optimization examples shown in Fig.~\ref{fig:techvalidation}. As seen, the optimized direction $D$ closely aligns with the actual optimization path $\Delta$. From the table in Fig.~\ref{fig:techvalidation}, it can be observed that the primary objective for the $A$ parameter combination is to optimize timing. The corresponding parameter set adjusts the work function to reduce the threshold voltage, increases the drive current by modifying \texttt{hfin}, \texttt{tfin}, and \texttt{lg}, and enhances both intra-cell and inter-cell routability by increasing \texttt{num\_rows}. These results align with physical expectations. Additionally, from the $B$ parameter combination, it is clear that the main objective is to reduce power. This leads to an opposite trend compared to the $A$ combination, where the work function is adjusted to increase the threshold voltage, reduce the drive current, and reduce \texttt{num\_rows} to minimize parasitic capacitance, thereby decreasing dynamic power. This analysis further validates the effectiveness of our proposed method.

% \begin{figure}
%     \centering
%     \includegraphics[width=1\linewidth]{figure/cosine_similarity.pdf}
%     \caption{tech validation.}
%     \label{fig:techvalidation}
% \end{figure}
\section{Conclusion}
\label{sec:conclusion}
This paper introduces \papername{}, a dual-loop automated framework for system-technology co-optimization (STCO). \papername{} combines system-level and technology-level optimizations through an interloop coordination mechanism, bridging the gap between system requirements and technology innovations while optimizing both levels simultaneously.
Evaluated on 7nm technology, \papername{} achieves \delayreduction{} delay reduction at iso-power and \powerreduction{} power savings at iso-delay compared to baseline approaches, complemented by a 33.2\% PPA hypervolume improvement that redefines Pareto optimality for cross-layer design.
Overall, \papername{} offers a promising solution to the challenges of scaling in the VLSI industry, providing a comprehensive and efficient methodology for STCO that can adapt to evolving technological demands. In the future, we aim to expand \papername{} to support a broader range of architectures and process technologies, further enhancing its versatility and impact in optimizing future VLSI designs.
 % \bk{add the future work extended to more architecture, technology }

% \input{tex/9-acknowledgment}

\clearpage

{
\bibliographystyle{IEEEtran}
\bibliography{./references}
}

\end{document}